\documentclass[sigconf, authorversion=True]{acmart}
\usepackage{multirow}
\usepackage{array} 
\usepackage{arydshln} 
\usepackage{bm}
\usepackage{mathtools}
\usepackage{diagbox}
\usepackage{url}

\AtBeginDocument{%
  }

\copyrightyear{2025}
\acmYear{2025}
\setcopyright{cc}
\setcctype{by}
\acmConference[ICMI '25]{Proceedings of the 27th International Conference on Multimodal Interaction}{October 13--17, 2025}{Canberra, ACT, Australia}
\acmBooktitle{Proceedings of the 27th International Conference on Multimodal Interaction (ICMI '25), October 13--17, 2025, Canberra, ACT, Australia}
\acmDOI{10.1145/3716553.3750764}
\acmISBN{979-8-4007-1499-3/2025/10}

\acmSubmissionID{1105}

\begin{document}

\title{Real-time Generation of Various Types of Nodding\\for Avatar Attentive Listening System}

\author{Kazushi Kato}
\affiliation{%
\institution{Kyoto University}
\country{Japan}}
\email{katou@sap.ist.i.kyoto-u.ac.jp}

\author{Koji Inoue}
\affiliation{%
\institution{Kyoto University}
\country{Japan}}
\email{inoue@sap.ist.i.kyoto-u.ac.jp}

\author{Divesh Lala}
\affiliation{%
\institution{Kyoto University}
\country{Japan}}
\email{lala@sap.ist.i.kyoto-u.ac.jp}

\author{Keiko Ochi}
\affiliation{%
\institution{Kyoto University}
\country{Japan}}
\email{ochi@sap.ist.i.kyoto-u.ac.jp}

\author{Tatsuya Kawahara}
\affiliation{%
\institution{Kyoto University}
\country{Japan}}
\email{kawahara@i.kyoto-u.ac.jp}
 
\renewcommand{\shortauthors}{Kato et al.}

\begin{abstract}
In human dialogue, nonverbal information such as nodding and facial expressions is as crucial as verbal information, and spoken dialogue systems are also expected to express such nonverbal behaviors. We focus on nodding, which is critical in an attentive listening system, and propose a model that predicts both its timing and type in real time. The proposed model builds on the voice activity projection (VAP) model, which predicts voice activity from both listener and speaker audio. We extend it to prediction of various types of nodding in a continuous and real-time manner unlike conventional models. In addition, the proposed model incorporates multi-task learning with verbal backchannel prediction and pretraining on general dialogue data. In the timing and type prediction task, the effectiveness of multi-task learning was significantly demonstrated. We confirmed that reducing the processing rate enables real-time operation without a substantial drop in accuracy, and integrated the model into an avatar attentive listening system. Subjective evaluations showed that it outperformed the conventional method, which always does nodding in sync with verbal backchannel. The code and trained models are available at \url{https://github.com/MaAI-Kyoto/MaAI}.
\end{abstract}




\maketitle

\section{Introduction}

In human dialogue, nonverbal information, such as nodding, eye contact, and facial expressions, plays as important a role as verbal information. 
In spoken dialogue systems and conversational robots, expressing such nonverbal information appropriately is expected to realize more natural interactions. 
For instance, in an attentive listening system~\cite{inoue2020attentive} that focuses on listening to the user, it is essential to express appropriate nonverbal listener responses in a timely manner.

Recently, research on generating more human-like nonverbal gestures has been actively conducted.
The GENEA Challenge~\cite{yoon2024} provides a common dataset for creating models of nonverbal gesture generation during dialogues, followed by the evaluation of the submitted models.
Most models contributed to this challenge generate speaker gestures from speech signals and real-time gesture generation models have been proposed~\cite{chen2024}. 
However, a few studies focusing on generating listener gestures were made~\cite{wolfert2023,schmuck2023}.

On the other hand, research on Listening Head Generation focuses on the listener’s nonverbal reactions in dialogue, with the aim of generating listener gestures based on the speaker’s gestures and speech signals.
This task is actively progressing, with many studies proposing generation models~\cite{dim2024,infp2024,difflistener2025}, and some achieving real-time generation~\cite{active2025}.
However, these models generate listener gestures in the same time frame as the given speaker's gestures and speech signals.
To achieve more natural and smooth interactions in spoken dialogue systems, it is necessary to predict listener gestures in future frames.

In response to this background, the Responsive Listening Head Generation~\cite{responsive2022} aims to predict the listener's gestures in the next frame based on the speaker's speech and gestures.
The benchmark dataset and baseline models have been provided, and several models have been proposed for this task~\cite{liu2024,tamon2024}.
They have identified challenges related to the natural timing of generated gestures~\cite{tamon2024}.

Nodding is a nonverbal behavior in dialogue in which participants shake their heads vertically~\cite{kondo2005}.
It may be performed simultaneously with verbal backchannel, such as "um" or "uh-huh", or it may be performed alone.
In dialogue, the listener's nodding plays a role in encouraging the speaker to continue speaking and is closely related to turn-taking~\cite{maynard1993}.
Several models have been proposed for predicting nodding, such as a model based on linear combinations of prosodic information~\cite{watanabe2000}, a model that predicts from given backchannel~\cite{mori2022_bc}, and a model that simultaneously predicts nodding and backchannel in multiparty dialogues~\cite{mori2022_mp}.

In this study, we propose a real-time prediction model that predicts the timing and type of nodding as a nonverbal listener response.
Assuming integration into an attentive listening dialogue system that focuses on listening to the user attentively and empathetically, the proposed model targets affirmative and responsive nods. 
To enable the prediction of nodding in future frames, each ground-truth label of nodding was offset 500 ms earlier than the transcript during training.
Using the proposed model, we could develop a system that can generate timely nodding in accordance with the speaker's speech, and appropriately encourage the speaker to speak and change speakers.
To the best of our knowledge, research on predicting various types of nodding in a continuous and real-time manner has not been made so far.
Since there is a relationship between the forms of nodding and backchannel that co-occur with each other~\cite{mori2021}, it is suggested that various types of nodding have different roles in dialogue.
Appropriately expressing various types of nodding in a spoken dialogue system should therefore yield more natural listener responses.

\begin{figure}[t]
  \begin{center}
  \includegraphics[width=\linewidth]{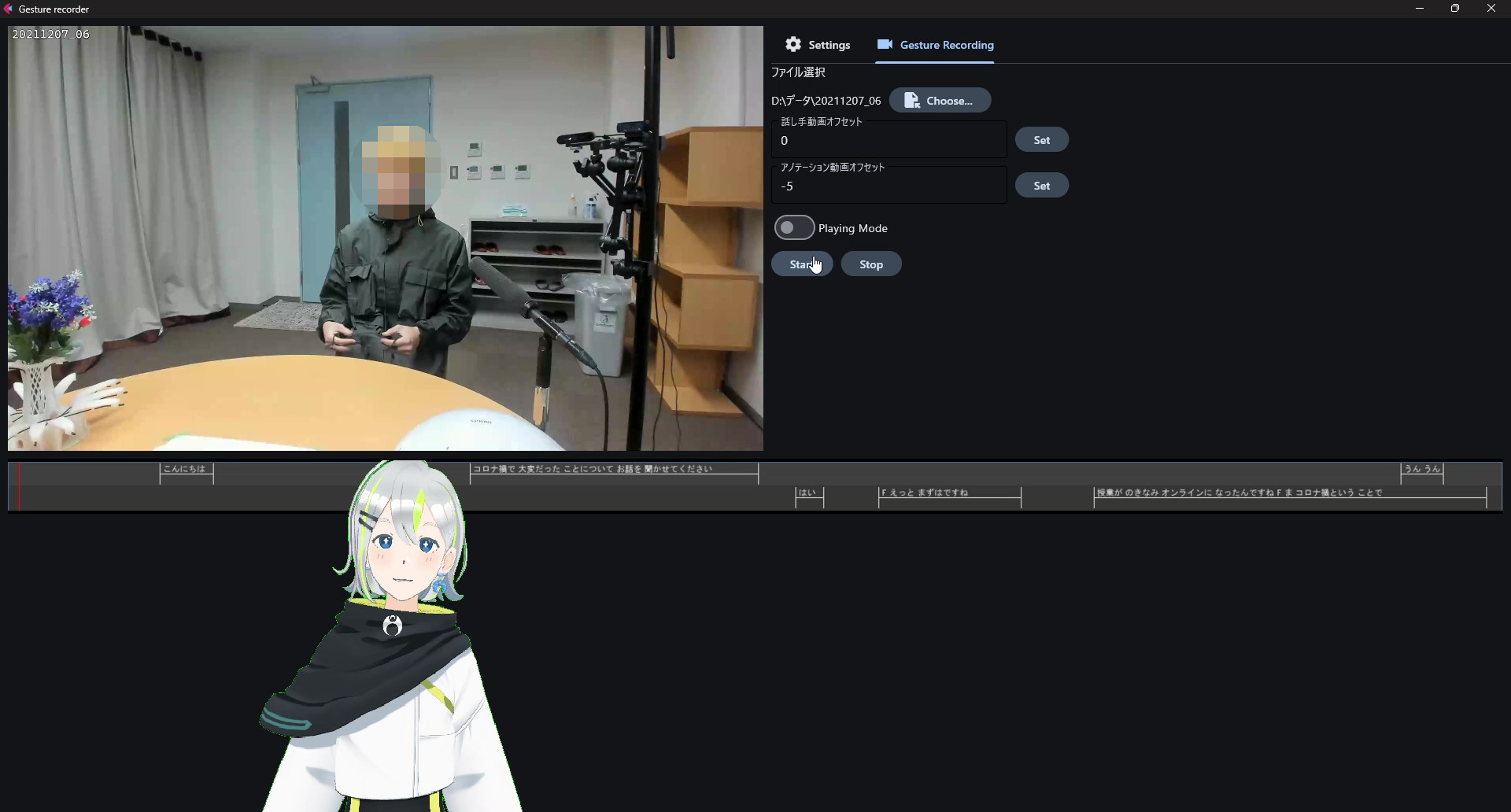}
  \caption{UI used for gesture recording}\label{fg:recording}
  \end{center}
\end{figure}

The proposed model is based on the architecture of Voice Activity Projection (VAP)~\cite{ekstedt2022}.
VAP predicts the participant's speech in future frames, and a model for application to verbal backchannel prediction has been proposed~\cite{inoue2025yeah}.
In this study, the speech prediction model is applied to nodding prediction, which has the following features.
First, it takes both the speaker's and the listener's speech signals as input to directly predict listener nodding in an end-to-end manner. 
Second, multi-task learning with backchannel prediction is conducted. 
Since backchannel and nodding are related in terms of occurrence timing and type~\cite{mori2020}, this multi-task learning is expected to improve nodding prediction accuracy.
An additional advantage of using the VAP model is that it can be pretrained with a general dialogue dataset consisting only of speech signals before finetuning with a smaller dataset containing visual nodding and vocal backchannel annotations.
Additionally, since VAP is lightweight and capable of real-time operation~\cite{inoue2024realtimeturntaking}, we evaluate the real-time processing performance of the proposed model, with a scope of its integration into an avatar attentive listening system.

\section{Dataset} \label{sec:data}

This study uses the attentive listening dataset collected through Wizard-of-Oz experiments with the android ERICA~\cite{kawahara2022}. 
This dataset was recorded where an operator (a trained actor) remotely controlled the robot using their own voice to engage in attentive listening with elderly people and university students. 
Since the data about nodding were not included, a nodding dataset was newly created by additionally recording listener gestures to the existing data.

\subsection{Recording of Listener Gestures} \label{sec:sec-gesture_record}

The same operator who controlled ERICA in the previous experiments reviewed the recorded dialogues and performed listener gestures, which were recorded. 
Webcam Motion Capture\footnote{\url{https://webcammotioncapture.info/index.php}} and MMDAgent-EX\footnote{\url{https://github.com/mmdagent-ex/MMDAgent-EX}} were used for gesture recording. 
Webcam Motion Capture captured motion data, including head movements, facial expressions, and blinking, from webcam video, and transmitted it to MMDAgent-EX. 
MMDAgent-EX recorded the received motion data while rendering it on a CG avatar, providing feedback to the operator during data collection (Figure \ref{fg:recording}). 
Listener gestures from 90 dialogues, averaging 8 minutes each, were recorded. 
This study used 72 dialogues for training, 9 for validation, and 9 for testing.

\subsection{Nodding Annotation} \label{subsec-nod_detect}

\begin{figure}[t]
    \begin{center}
    \includegraphics[width=\linewidth]{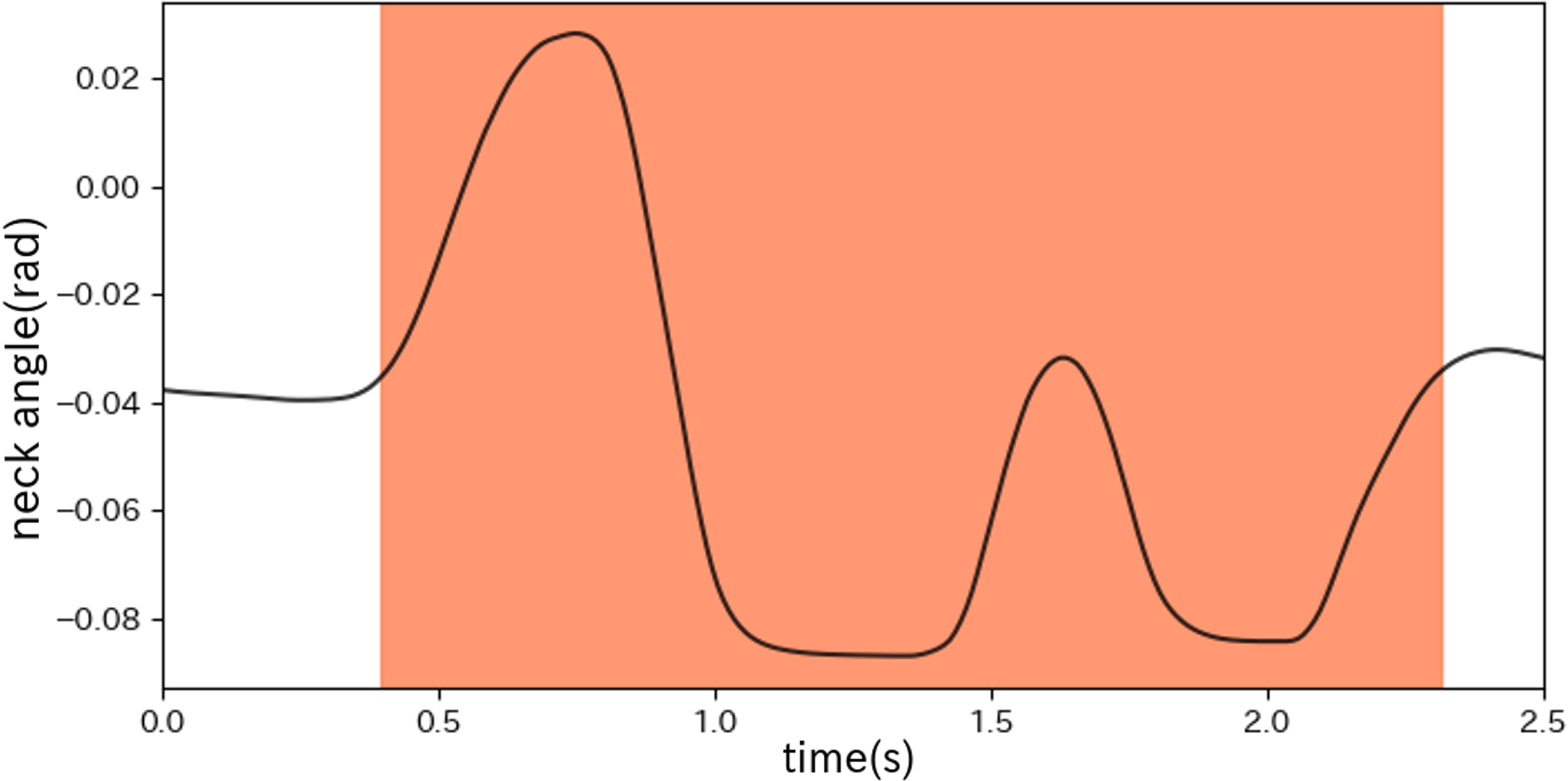}
    \caption{Smoothed motion data and detected nodding segment}\label{fg:recorded_nod_detect}
    \end{center}
\end{figure}

To detect nodding events from the recorded motion data, we analyzed data corresponding to the vertical neck angle (radians). 
The data were downsampled to 100 Hz, smoothed with a moving average of 7 frames, and nodding segments were detected based on the gradient of the smoothed data (Figure \ref{fg:recorded_nod_detect}).

{\tabcolsep = 0.35cm
  \begin{table}[t]
    \begin{center}
      \caption{Distribution of nodding types}\label{tb:nod}
      \begin{tabular}{lccc} \hline
        Types&Time(s)&Ratio(\%)&Counts \\ \hline
        short&\phantom{2}3502.4&\phantom{2}8.9&4227 \\
        long&\phantom{2}4883.5&12.4&3446 \\
        long\_p&\phantom{2}1762.0&\phantom{2}4.4&1008 \\
        no nodding&29092.0&74.1& \\ \hline
      \end{tabular}
    \end{center}
  \end{table}
}

\begin{figure*}[t]
    \begin{center}
    \includegraphics[width=\linewidth]{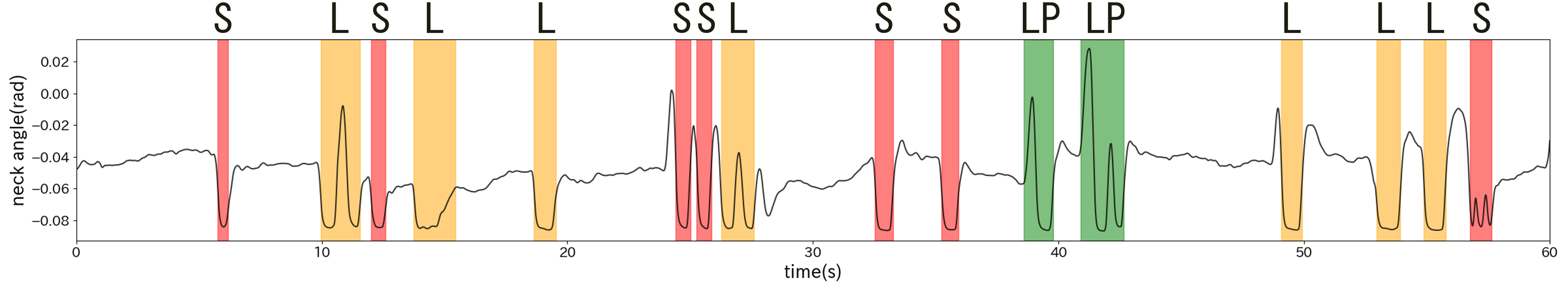}
    \caption{Examples of nodding annotation (red: short (S), yellow: long (L), green: long\_p (LP))}\label{fg:sample_annotation}
    \end{center}
\end{figure*}

According to previous studies, nodding co-occurring with continuer backchannel has a smaller average range of movement, whereas that co-occurring with assessment backchannel and lexical responses has a larger average range of movement~\cite{mori2021}.
Therefore, nodding with small and large movement ranges may have different functions and meanings during dialogue.
In addition, nodding with swinging up is regarded to reflect a cognitive shift in the listener, and is more likely to co-occur with assessment backchannel than with continuer backchannel~\cite{mori2020}.
Accordingly, predicting whether or not a nod includes swinging up is necessary for expressing a cognitive shift in the listener.
Referring to those previous studies, we annotated three types of nodding for each detected nodding segment (Figure \ref{fg:sample_annotation}):
\begin{itemize}
  \item short: Small movement range, regardless of swinging up.
  \item long: Large movement range without swinging up.
  \item long\_p: Large movement range with swinging up.
\end{itemize}
The distribution of the three nodding types is presented in Table \ref{tb:nod}. 
Nodding segments accounted for 25.9\% of the total time, with short, long, and long\_p occurring at 8.9\%, 12.4\%, and 4.4\%, respectively.

\section{Proposed Model} \label{sec:proposed}

This section provides a description of the proposed model for achieving continuous and real-time nodding prediction.
We first describe our proposed model architecture, followed by multi-task learning with verbal backchannel prediction and finetuning for nodding prediction.

\subsection{Architecture} \label{sec:VAP}

The architecture of the proposed model based on VAP is shown in Figure \ref{fg:VAP_ST}. 
In the original VAP, the audio waveforms of the two dialogue participants are respectively encoded using contrastive predictive coding (CPC), then processed by Self-attention Transformers. 
These representations are then passed through Cross-attention Transformers, which allow references to each other's attention states.
Finally, task-specific linear layers produce the outputs for voice activity detection (VAD) and voice activity projection (VAP).
VAD means detection of voice activity in the current input speech frames and VAP means prediction of voice activity for the next two seconds, which corresponds to the implicit prediction of a turn transition occurring within the next two seconds.

We extend the VAP model by introducing an additional linear layer for nodding prediction. 
The loss function is defined as Equation \eqref{eq:VAP_loss_nod}.

\begin{gather}
L=L_{nod}+w_{vad} L_{vad}+w_{vap} L_{vap}\label{eq:VAP_loss_nod}
\end{gather}
$L_{vad}$ and $L_{vap}$ represent the loss of voice activity detection (VAD) and voice activity projection (VAP) in the original VAP model. 
The weights $w_{vad}$ and $w_{vap}$ are hyperparameters used to adjust the weighting of the loss terms.
$L_{nod}$ represents the cross-entropy loss of nodding prediction and is defined by Equation \eqref{eq:loss_nod}.
\begin{gather}
  L_{nod}=-\sum_{c}^{C} {\bm{r}_{nod}}^{(c)} \log{{\bm{o}_{nod}}^{(c)}} \label{eq:loss_nod}
\end{gather}
where $C$ is the number of nodding-type classes, $\bm{o}_{nod}(\in [0,1] ^C)$ is the predicted probability of nodding converted from the output of the linear layer, and $\bm{r}_{nod}(\in \{0,1\}^C)$ is the one-hot vector of the ground-truth label.

\subsection{Multi-task Learning with Backchannel Prediction}

\begin{figure}[t]
  \begin{center}
      \includegraphics[width=\linewidth]{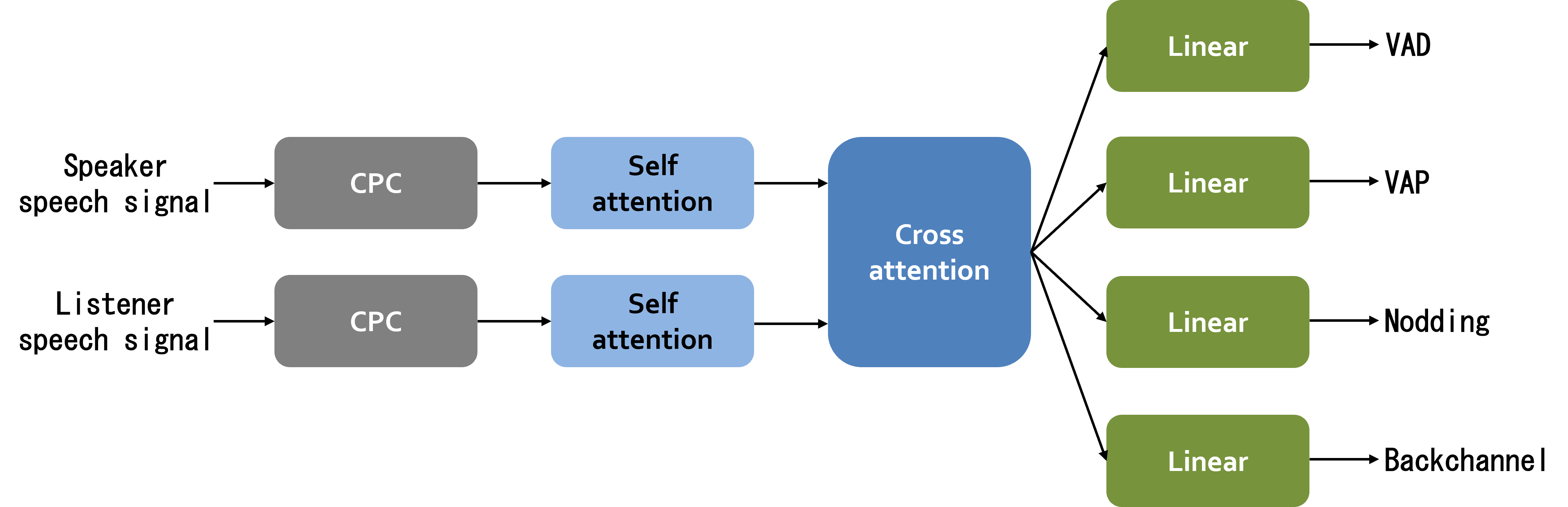}
      \caption{Proposed model: multi-task learning with VAP and VAD and backchannel prediction}\label{fg:VAP_MT}
  \end{center}
\end{figure}

In the proposed model, multi-task learning with verbal backchannel prediction is performed.
For the multi-task learning, a linear layer for backchannel prediction is added after the Cross-attention Transformer layer (Figure \ref{fg:VAP_MT}). 
The loss for backchannel prediction $L_{bc}$ is defined as cross-entropy loss in the same way of Equation \eqref{eq:loss_nod}. 
This loss is multiplied by the weight $w_{bc}$ and added to the total loss (Equation \eqref{eq:VAP_loss_nod_bc}).

\begin{gather}
L=L_{nod}+w_{vad} L_{vad}+w_{vap} L_{vap}+w_{bc} L_{bc}\label{eq:VAP_loss_nod_bc}
\end{gather}

In the proposed model, only the timing of the backchannel is predicted. Considering that backchannel and nodding often occur simultaneously, self-feedback of the predicted results of backchannel prediction to the listener's speech signal input is expected to suppress the continuous prediction of nodding.

\subsection{Finetuning for Nodding Prediction}

\begin{figure}[t]
  \begin{center}
      \includegraphics[width=\linewidth]{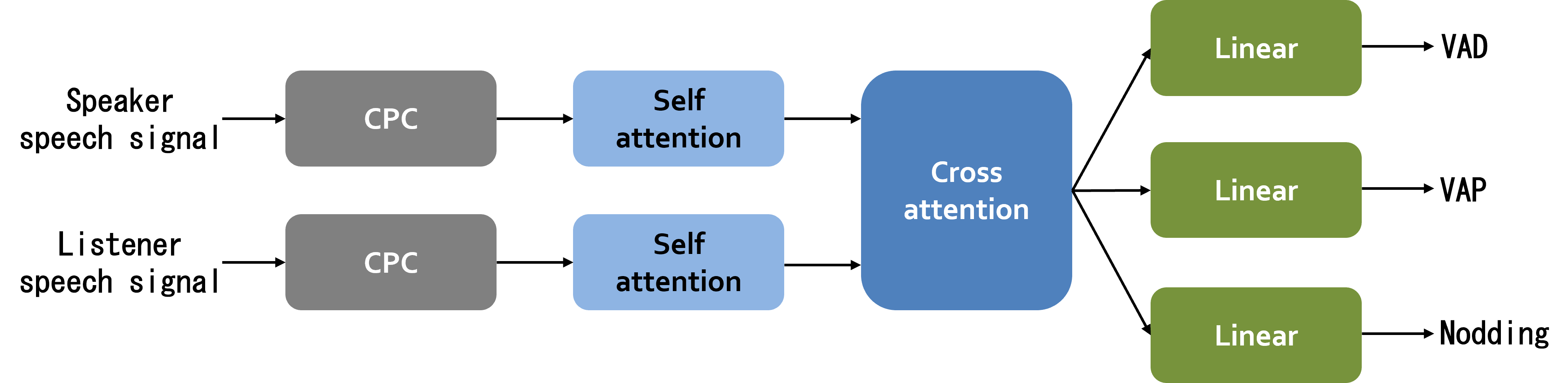}
      \caption{Proposed model: multi-task learning with VAP and VAD}\label{fg:VAP_ST}
  \end{center}
\end{figure}

With a large amount of general dialogue dataset, the Self-attention layer and Cross-attention layer in Figure \ref{fg:VAP_ST} can be pretrained through the voice activity detection and projection task with the original VAP loss function.
After this pretraining, the model is finetuned with a dataset specialized for backchannel and nodding prediction with the loss function \eqref{eq:VAP_loss_nod} or \eqref{eq:VAP_loss_nod_bc}.
We evaluated whether the pretraining enhances nodding prediction accuracy.

\section{Experimental Evaluation} \label{sec:experiment}
The proposed model was evaluated in the following two tasks for nodding prediction: the timing prediction task and the timing and nodding type prediction task at each frame.
The evaluation metrics were F1-score (F1), precision (Pre.), and recall (Rec.) calculated at the frame level.

For pretraining the model, 203 dialogues of Wizard-of-Oz data collected with ERICA are used.
This includes 72 dialogues used as data for training described in Section \ref{sec:data}.
In addition to attentive listening, the dataset contains two tasks of job interviews and first-meeting conversations.

The CPC component is pretrained on approximately 60,000 hours of Librispeech data, and its parameters are frozen during training. 
The hyperparameters $w_{vad}$, $w_{vap}$ and $w_{bc}$ were respectively set to 0.2, 0.2 and 0.5 with a preliminary experiment.

As a comparative model, we also tested a monaural model that takes only the speaker’s speech signal as input. 
Similarly to the proposed model, CPC encodes the speech signal, followed by a self-attention layer and task-specific linear layers.

\subsection{Timing Prediction} \label{sec:experiment_timing}

\begin{figure}[t]
    \begin{center}
    \includegraphics[width=\linewidth]{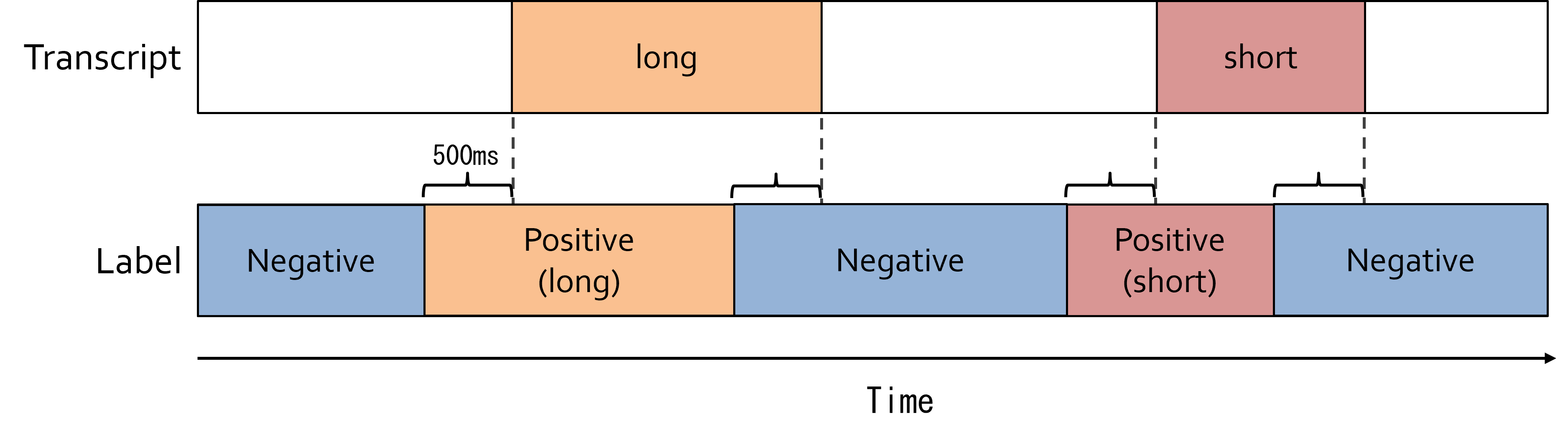}
    \caption{Definition of ground-truth labels}\label{fg:label}
    \end{center}
\end{figure}

The first task is to predict whether or not nodding occurs at each frame.
To account for the processing delay from the time the model predicts nodding until it is actually executed, we offset the ground-truth nodding interval 500 ms earlier, as illustrated in Figure \ref{fg:label}. 
This means that the model predicts the probability that a nodding will occur after 500 ms.
When calculating the loss for backchannel and nodding, we addressed the ratio of positive and negative samples by assigning a weight to positive samples that is triple the weight of negative samples.

The following models were evaluated in the experiment.
\begin{itemize}
    \setlength{\itemsep}{0pt}
	\setlength{\parskip}{0pt}
    \item (Random) Always predicting nodding in all frames.
    \item Monaural model (Mono)
        \begin{itemize}
            \item (ST) Only learning nodding prediction.
            \item (MT w/ BC) Multi-task learning with verbal backchannel prediction.
        \end{itemize}
    \item Proposed model (Proposed)
        \begin{itemize}
            \item (ST) Only learning nodding prediction in the proposed model.
            \item (ST w/ PT) Pretraining VAP in addition to the above model.
            \item (MT w/ BC) Multi-task learning with verbal backchannel prediction in the proposed model.
            \item (MT w/ BC, PT) Pretraining VAP in addition to the above model.
        \end{itemize}
\end{itemize}

The results are presented in Table \ref{tb:result_timing}.
Both the single-task (ST) and multi-task with backchannel (MT w/ BC) of the proposed model demonstrate a higher score than the Random and monaural models in terms of F1-score.
Moreover, VAP pretraining shows some effect in both single-task and multi-task models.
In contrast, multi-task learning with verbal backchannel has little synergy effect.

We also conducted statistical tests to determine whether differences in F1 scores between these models were statistically significant.
Statistical tests were conducted using bootstrap resampling on the test data, with 1,000 iterations.
One-tailed \textit{t}-tests were performed to evaluate whether the F1-score of the one model was significantly higher than that of the other model. 
The results are shown in Table \ref{tb:result_bootstrap_timing}.
Compared to the monaural model, all proposed models showed statistically significant improvement in F1 score.
However, the multi-task learning model did not show a statistically significant improvement in F1-score over the single-task model. 
Similarly, the pretrained models did not achieve significantly higher F1 scores than models without pretraining.

\begin{figure*}[t]
  \begin{center}
  \includegraphics[width=\linewidth]{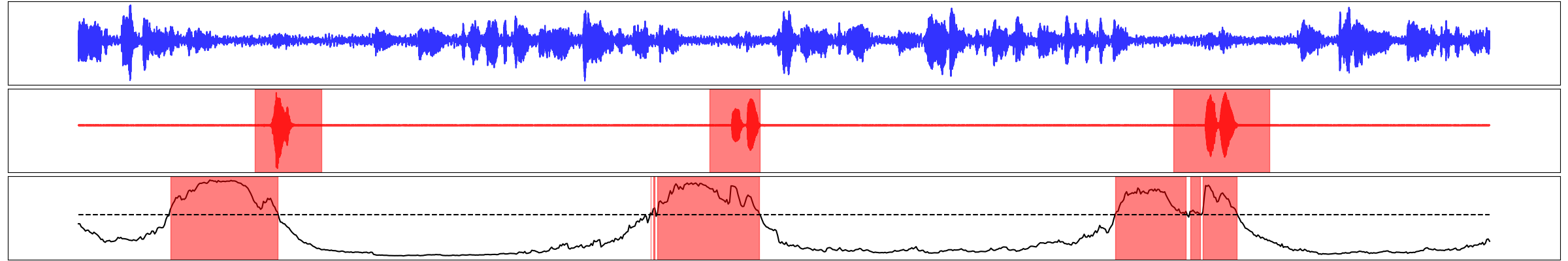}
  \caption{Output samples of timing prediction model (The panels, from the top to bottom, represent speaker speech signal, listener speech signal with nodding segments, probability of nodding occurrence with predicted segments, respectively.)}\label{fg:VAP_timing_result}
  \end{center}
\end{figure*}

Figure \ref{fg:VAP_timing_result} shows sample outputs from the MT w/ BC model, showing that it predicts earlier than their actual occurrence.
It was observed that the predicted probability of nodding decreases when the avatar is speaking. 
This suggests that by simultaneously predicting backchannel and feeding the prediction results back into the model, excessive nodding can be suppressed.

\subsection{Timing and Type Prediction} \label{sec:experiment_timing_kinds}

\begin{table}[t]
  \begin{center}
    \caption{Results for timing prediction}\label{tb:result_timing}
    \begin{tabular}{llccc} \hline
      \multicolumn{2}{c}{Method}&F1&Pre.&Rec. \\ \hline
      \multicolumn{2}{c}{Random}&37.20&22.85&100.00 \\ \hdashline
      \multirow{2}{*}{Mono}&ST&48.61&44.03&\phantom{1}54.26 \\
      &MT w/ BC&49.16&42.53&\phantom{1}58.25 \\ \hdashline
      \multirow{4}{*}{Proposed}
      &ST&55.47&46.91&\phantom{1}67.85 \\
      &ST w/ PT&55.92&\textbf{47.27}&\phantom{1}68.44 \\
      &MT w/ BC&55.89&47.01&\phantom{1}68.90 \\
      &MT w/ BC, PT&\textbf{55.93}&46.64&\phantom{1}\textbf{69.82} \\ \hline
    \end{tabular}
  \end{center}
\end{table}

\begin{table}[t]
  \begin{center}
    \caption{\textit{t}-test results for F1-score differences in timing prediction task}\label{tb:result_bootstrap_timing}
    \begin{tabular}{cr@{\ vs.\ }r r} \hline
        \multicolumn{3}{c}{Comparisons}& p-value\phantom{*} \\ \hline
        \multirow{5}{*}{Proposed vs. Mono} & \multicolumn{1}{c}{Proposed\phantom{**}} & \multicolumn{1}{c}{Mono\phantom{**}} & \\ \cline{2-3}
        & ST & ST & <.001** \\
        & ST w/ PT & ST & <.001** \\
        & MT w/ BC & MT w/ BC & <.001** \\
        & MT w/ BC, PT & MT w/ BC & <.001** \\ \hdashline
        \multirow{3}{*}{MT vs. ST} & \multicolumn{2}{c}{Proposed\phantom{**}} \\ \cline{2-3}
        & MT w/ BC & ST & \phantom{<}.074\phantom{**} \\
        & MT w/ BC, PT & ST w/ PT & \phantom{<}.483\phantom{**} \\ \hdashline
        \multirow{2}{*}{w/ PT vs. w/o PT} & ST w/ PT & ST & \phantom{<}.130\phantom{**} \\
        & MT w/ BC, PT & MT w/ BC & \phantom{<}.454\phantom{**} \\ \hline
        \multicolumn{4}{r}{**p<0.01} \\
    \end{tabular}
  \end{center}
\end{table}

The next task is to predict not only timing but also the type of nodding, or classify each frame into four classes (short, long, long\_p, and no-nodding).
Similar to Section \ref{sec:experiment_timing}, we offset the ground-truth nodding interval 500 ms earlier. 
We assigned the loss weight to positive samples five times larger than that of negative samples.

The results are presented in Table \ref{tb:result_timing_kinds_short}, Table \ref{tb:result_timing_kinds_long} and Table \ref{tb:result_timing_kinds_long_p}.
In prediction of short nodding and long nodding, both ST and MT w/ BC of the proposed model outperformed Random and the monaural model in F1-score.
In case of long\_p nodding, ST without pretraining scored lower than monaural model, but other proposed models outperformed it. 
In prediction of all types, the effectiveness of VAP pretraining was confirmed for both ST and MT w/ BC. 
Comparing ST and MT w/ BC, MT w/ BC outperformed ST in long nodding and lonp\_p nodding prediction but not in short nodding predicton.
This result indicates that long and long\_p nodding are more associated with verbal backchannel than short nodding. 

\begin{table}[t]
    \begin{center}
      \caption{Results for timing and type prediction (short)}\label{tb:result_timing_kinds_short}
      \begin{tabular}{llccc} \hline
        \multicolumn{2}{c}{Method}&F1&Pre.&Rec. \\ \hline
        \multicolumn{2}{c}{Random}&14.34&\phantom{1}7.72&100.00 \\ \hdashline
        \multirow{2}{*}{Mono}&ST&19.64&18.65&\phantom{1}20.73 \\
        &MT w/ BC&22.64&21.78&\phantom{1}23.57 \\ \hdashline
        \multirow{4}{*}{Proposed}&ST&29.19&21.07&\phantom{1}47.47 \\
        &ST w/ PT&\textbf{29.94}&21.04&\phantom{1}\textbf{51.91} \\
        &MT w/ BC&28.58&22.18&\phantom{1}40.15 \\
        &MT w/ BC, PT&28.86&\textbf{25.42}&\phantom{1}33.37 \\ \hline
      \end{tabular}
    \end{center}
\end{table}
\begin{table}[t]
    \begin{center}
      \caption{Results for timing and type prediction (long)}\label{tb:result_timing_kinds_long}
      \begin{tabular}{llccc} \hline
        \multicolumn{2}{c}{Method}&F1&Pre.&Rec. \\ \hline
        \multicolumn{2}{c}{Random}&21.71&12.18&100.00 \\ \hdashline
        \multirow{2}{*}{Mono}&ST&34.04&26.54&\phantom{1}47.47 \\
        &MT w/ BC&33.82&27.89&\phantom{1}42.94 \\ \hdashline
        \multirow{4}{*}{Proposed}&ST&36.06&\textbf{37.58}&\phantom{1}34.65 \\
        &ST w/ PT&36.07&36.94&\phantom{1}35.24 \\
        &MT w/ BC&38.69&28.28&\phantom{1}61.24 \\
        &MT w/ BC, PT&\textbf{39.17}&28.51&\phantom{1}\textbf{62.57} \\ \hline
      \end{tabular}
        \end{center}
\end{table}
\begin{table}[t]
    \begin{center}
      \caption{Results for timing and type prediction (long\_p)}\label{tb:result_timing_kinds_long_p}
      \begin{tabular}{llccc} \hline
        \multicolumn{2}{c}{Method}&F1&Pre.&Rec. \\ \hline
        \multicolumn{2}{c}{Random}&\phantom{1}5.64&\phantom{1}2.90&100.00 \\ \hdashline
        \multirow{2}{*}{Mono}&ST&19.10&16.13&\phantom{1}23.39 \\
        &MT w/ BC&17.46&15.54&\phantom{1}19.92 \\ \hdashline
        \multirow{4}{*}{Proposed}&ST&18.12&14.20&\phantom{1}25.03 \\
        &ST w/ PT&19.70&14.95&\phantom{1}28.90 \\
        &MT w/ BC&19.97&14.81&\phantom{1}\textbf{30.65} \\
        &MT w/ BC, PT&\textbf{22.09}&\textbf{18.95}&\phantom{1}26.46 \\ \hline
      \end{tabular}
    \end{center}
\end{table}

\begin{figure*}[t]
  \begin{center}
  \includegraphics[width=\linewidth]{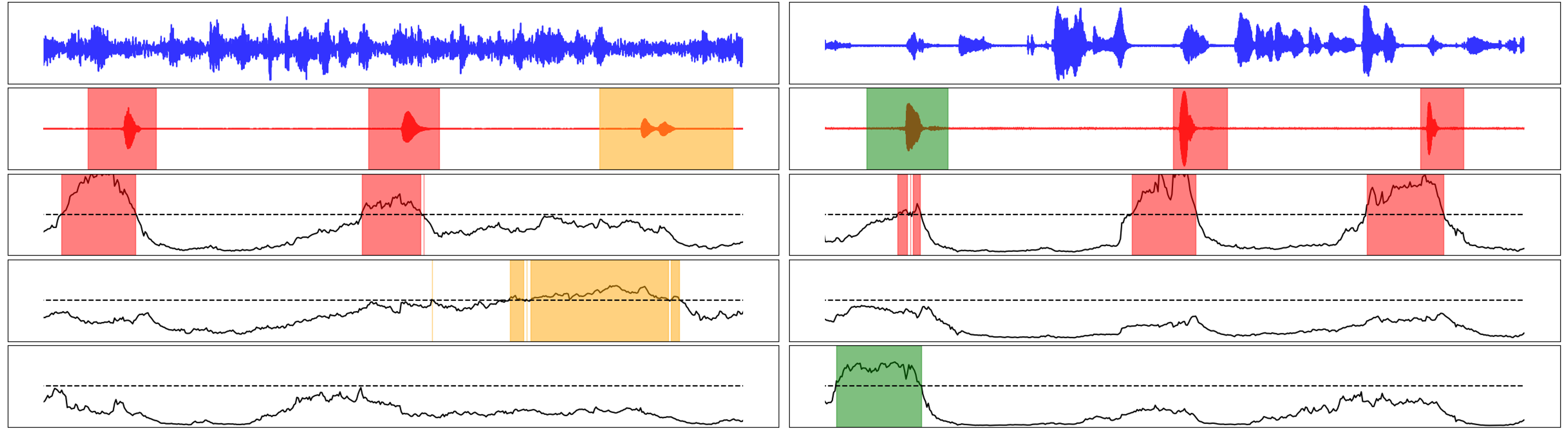}
  \caption{Output samples of timing and type prediction model (The panels, from the top to bottom, represent speaker speech signal, listener speech signal with nodding segments, probability of nodding occurrence with predicted segments (short, long, long\_p), respectively.)}\label{fg:VAP_timing_kinds_result}
  \end{center}
\end{figure*}

The equally-weighted average results are presented in Table \ref{tb:result_average}.
These results demonstrate that the multi-task learning with backchannel is effective. 
This is because backchannel and nodding often co-occur with each other and have the similar role of conveying to the speaker that the listener is listening to the speaker.
This result is consistent with previous studies on nodding behavior~\cite{mori2022_mp}. 
Additionally, VAP pretraining with general dialogue data improves performance. 
This suggests that the occurrence of nodding, a nonverbal response, may be related to speech activity, and that the versatility of the VAP is also effective in predicting nodding.

As in Section \ref{sec:experiment_timing}, we also conducted statistical tests to determine whether differences in averaged F1 scores between models were statistically significant. The results for the statistical tests are shown in Table \ref{tb:result_bootstrap_timing_kinds}. 
Compared to the monaural models, all proposed models showed statistically significant improvement in F1 score.
Moreover, unlike in the timing prediction task, the multi-task learning model achieved a significantly higher F1 score than the single-task model in the timing and type prediction task.
This result suggests that multi-task learning with backchannel prediction is particularly effective when predicting not only timing but also the type of nodding.

Figure \ref{fg:VAP_timing_kinds_result} shows sample output from the MT w/ BC, PT model, showing that it predicts types of nodding earlier than the actual occurrence.
As in Section \ref{sec:experiment_timing}, we observed that the prediction probability decreases when listener speech is present in the input.

\begin{table}[t]
    \begin{center}
      \caption{Equally-averaged results of short, long and long\_p}
      \label{tb:result_average}
      \begin{tabular}{llccc} \hline
        \multicolumn{2}{c}{Method} & F1 & Pre. & Rec. \\ \hline
        \multicolumn{2}{c}{Random}
          & 13.89 & \phantom{0}7.59 & 100.00 \\ \hdashline
        \multirow{2}{*}{Mono}
          & ST          & 24.26 & 20.43 & \phantom{1}30.53 \\
          & MT w/ BC    & 24.64 & 21.73 & \phantom{1}28.81 \\ \hdashline
        \multirow{4}{*}{Proposed}
          & ST          & 27.78 & 24.28 & \phantom{1}35.71 \\
          & ST w/ PT    & 28.57 & 24.31 & \phantom{1}38.68 \\
          & MT w/ BC    & 29.08 & 21.75 & \phantom{1}\textbf{44.01} \\
          & MT w/ BC, PT& \textbf{30.04} & \textbf{24.30} & \phantom{1}40.81 \\ \hline
      \end{tabular}
    \end{center}
\end{table}

\begin{table}[t]
  \begin{center}
    \caption{\textit{t}-test results for averaged F1-score differences in timing and type prediction}\label{tb:result_bootstrap_timing_kinds}
    \begin{tabular}{cr@{\ vs.\ }r c} \hline
        \multicolumn{3}{c}{Comparisons}& p-value\phantom{**} \\ \hline
        \multirow{5}{*}{Proposed vs. Mono} & \multicolumn{1}{c}{Proposed\phantom{**}} & \multicolumn{1}{c}{Mono\phantom{**}} & \\ \cline{2-3}
        & ST & ST & <.001** \\
        & ST w/ PT & ST & <.001** \\
        & MT w/ BC & MT w/ BC & <.001** \\
        & MT w/ BC, PT & MT w/ BC & <.001** \\ \hdashline
        \multirow{3}{*}{MT vs. ST} & \multicolumn{2}{c}{Proposed\phantom{**}} \\ \cline{2-3}
        & MT w/ BC & ST & \phantom{<}.049*\phantom{*} \\
        & MT w/ BC, PT & ST w/ PT & \phantom{<}.012*\phantom{*} \\ \hdashline
        \multirow{2}{*}{w/ PT vs. w/o PT} & ST w/ PT & ST & \phantom{<}.138\phantom{**} \\
        & MT w/ BC, PT & MT w/ BC & \phantom{<}.111\phantom{**} \\ \hline
        \multicolumn{4}{r}{*p<0.05  **p<0.01} \\
    \end{tabular}
  \end{center}
\end{table}

\subsection{Real-time Processing Performance} \label{sec:exp:realtime}
\begin{figure*}[t]
  \begin{center}
  \includegraphics[width=\linewidth]{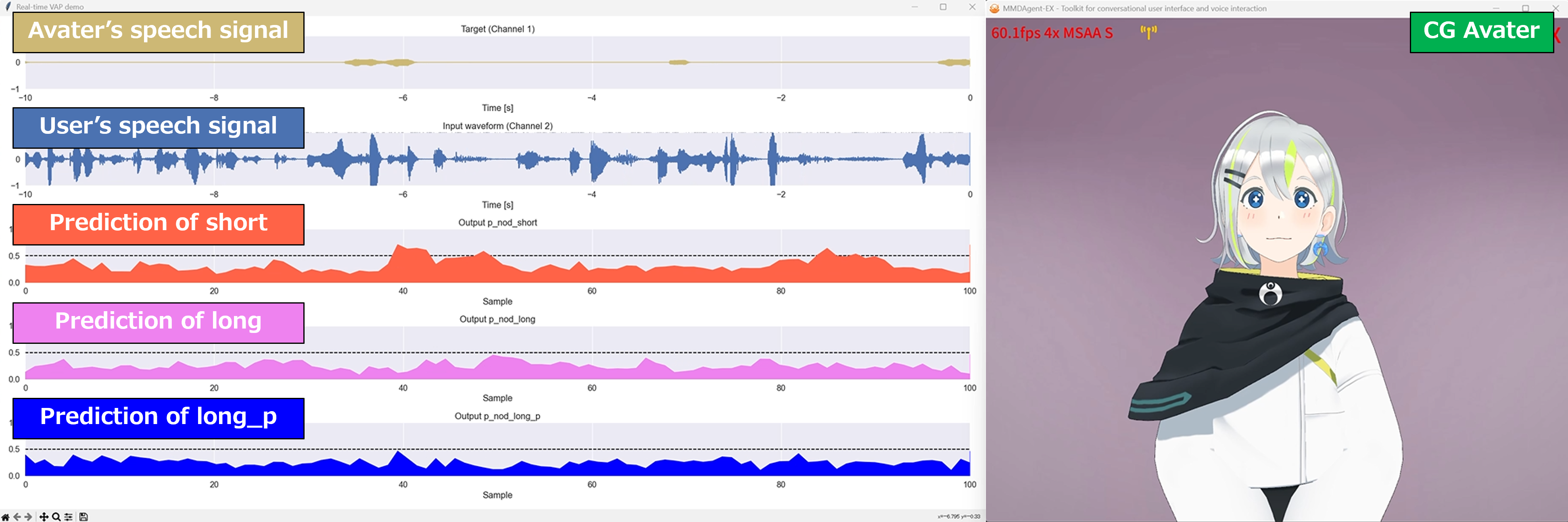}
  \caption{Avatar attentive listening system integrated with the proposed model}\label{fg:system}
  \end{center}
\end{figure*}

We also investigated the real-time processing performance of the proposed model.
The model evaluated was MT w/ BC, PT, which achieved the highest F1-score on average in timing and type prediction.
In both Section \ref{sec:experiment_timing} and Section  \ref{sec:experiment_timing_kinds}, processing frame rate and input signal length were set to 50 Hz and 20.0 s respectively.
We retrained the models for frame rates of 50 Hz and 10 Hz with input signal lengths of 20.0 s, 10.0 s, 5.0 s, 2.5 s, and 1.0 s, and measured F1-score and Real-time factor (RTF) on the test data.
We used only a CPU (Intel Core i5-14500 @ 2.60 GHz), assuming deployment on laptop-class devices without GPUs.

The results are shown in Table \ref{tb:result_real-time_50} and Table \ref{tb:result_real-time_10}. 
In both models, as the input length decreases, the F1-score tends to decline. 
However, the 10 Hz model, which reduced the processing frame rate to one-fifth from the 50 Hz model, had no substantial decline in F1-score.
In the 50 Hz model, the RTF exceeded 1.0 for all input lengths, whereas in the 10 Hz model, they all remained below 1.0.

{\tabcolsep = 0.35cm
\begin{table}[t]
  \begin{center}
    \caption{F1-score and Real-time factor for each input length (50 Hz model)}\label{tb:result_real-time_50}
    \begin{tabular}{ccccc} \hline
      \multirow{2}{*}{Input length(s)}&\multicolumn{3}{c}{F1-score}&\multirow{2}{*}{RTF} \\ \cline{2-4}
      &short&long&long\_p& \\ \hline
      20.0&28.87&39.17&22.10&3.55 \\
      10.0&27.27&38.78&20.24&2.91 \\
      \phantom{1}5.0&26.33&37.33&19.12&2.67 \\
      \phantom{1}2.5&25.32&35.81&19.58&2.67 \\
      \phantom{1}1.0&25.19&34.10&17.84&2.72 \\ \hline
    \end{tabular}
  \end{center}
\end{table}
\begin{table}[t]
    \begin{center}
    \caption{F1-score and Real-time factor for each input length (10 Hz model)}\label{tb:result_real-time_10}
    \begin{tabular}{ccccc} \hline
      \multirow{2}{*}{Input length(s)}&\multicolumn{3}{c}{F1-score}&\multirow{2}{*}{RTF} \\ \cline{2-4}
      &short&long&long\_p& \\ \hline
      20.0&29.43&37.74&20.82&0.49 \\
      10.0&29.08&37.49&20.46&0.47 \\
      \phantom{1}5.0&27.62&37.31&22.52&0.45 \\
      \phantom{1}2.5&26.95&36.23&20.85&0.46 \\
      \phantom{1}1.0&24.98&34.68&15.72&0.49 \\ \hline
    \end{tabular}
  \end{center}
\end{table}
}

These results suggest that reducing the frame rate to 10 Hz does not significantly degrade accuracy. 
When integrating the proposed model into a spoken dialogue system, using the 10 Hz model with an input length of around 10.0 s is considered appropriate.

\section{Subjective Evaluation}
We integrated the proposed model into our avatar attentive listening system with CG agent\footnote{CG-CA Gene (c) 2023 by Nagoya Institute of Technology, Moonshot R\&D Goal 1 Avatar Symbiotic Society} (Figure \ref{fg:system}). 
In this system, three types of nodding, as described before, can be expressed in real time in response to the user's utterances. 
The nodding motions were newly created specifically for the system. 
We conducted a subjective evaluation to investigate whether the proposed nodding prediction model gives a better impression to users compared to conventional methods.

\subsection{Method}
The evaluation was conducted for the following four methods.
\begin{itemize}
    \setlength{\itemsep}{0pt}
	\setlength{\parskip}{0pt}
    \item Conventional methods
        \begin{itemize}
            \item (Only BC) Predicting and expressing verbal backchannels only.
            \item (ND w/ BC) Expressing a single type of nodding along with predicted backchannel.
        \end{itemize}
    \item Proposed methods
        \begin{itemize}
            \item (Only ND) Predicting and expressing three types of nodding only.
            \item (BC \& ND) Predicting and expressing backchannels and three types of nodding, respectively.
        \end{itemize}
\end{itemize}
For backchannel prediction, a VAP-based model was used~\cite{inoue2025yeah}.
For the experiment, we first conducted 1–2 minute attentive listening using the system integrated with the proposed method (BC \& ND) and recorded the audio.
This procedure was repeated nine times, resulting in nine different audio recordings. 
Each of these recordings was then streamed into the avatar attentive listening system implemented with each of the four methods described above. 
For each case, a video was recorded showing the avatar expressing backchannel or nodding in response to the audio.
In total, 36 demonstration videos were produced.

The nine audio recordings were grouped into three sets, each consisting of three recordings, and fifteen different crowdsource workers were recruited for each set. 
In total, 45 participants took part in the subjective evaluation experiment.
Each worker in each set watched twelve videos in total (three audio recordings $\ast$ four methods) and evaluated each video based on the following four metrics.
\begin{itemize}
  \item How human-like is the avatar's response? (\textbf{human likeness})
  \item How natural is the avatar's response? (\textbf{naturalness})
  \item How well does the avatar appear to be listening to the user? (\textbf{attentiveness})
  \item How well does the avatar encourage the user to continue speaking? (\textbf{facilitation})
\end{itemize}
Each metric was rated on a 7-point scale from 1 to 7.
The videos were presented to the workers in a randomized order.
This experiment was conducted in Japanese.

\subsection{Results and Analysis}
\begin{table}[t]
  \begin{center}
    \caption{Averaged scores of evaluation on subjective experiment (Hum : human likeness, Nat : naturalness, Att : attentiveness, Fac : facilitation) (n=45)}\label{tb:result_subjective_evaluation}
    \begin{tabular}{ccccc} \hline
      &\multicolumn{2}{c}{Conventional methods}&\multicolumn{2}{c}{Proposed methods} \\ \hline
      &Only BC&ND w/ BC&Only ND&BC \& ND \\ \hline
      Hum&3.76&5.00&4.48&\textbf{5.14} \\
      Nat&3.57&4.39&4.57&\textbf{4.64} \\
      Att&4.08&4.87&4.76&\textbf{5.34} \\ 
      Fac&3.69&4.38&3.60&\textbf{4.68} \\ \hline
    \end{tabular}
  \end{center}
\end{table}
The results for the subjective evaluation are shown in Table \ref{tb:result_subjective_evaluation}. 
Each worker viewed three videos per method and their ratings were averaged to obtain a single evaluation score per method per worker.
These scores were then averaged across workers for each combination of the evaluation criteria and methods. 

Across all evaluation metrics, the method that predicts backchannel and nodding using separate models achieved the highest scores. 
Comparison of ND w/ BC and BC \& ND reveals a particularly large difference in \textbf{attentiveness} among all metrics, suggesting that predicting various types of nodding can make users feel that the dialogue system is listening attentively to their speech.
Additionally, a comparison of Only BC with ND w/ BC or BC \& ND shows that systems expressing both backchannel and nodding achieve higher overall scores than those using only backchannel. 
This suggests that the nonverbal listener response of nodding enhances the system’s human-likeness and perceived attentiveness.

\begin{table}[t]
  \begin{center}
    \caption{\textit{t}-test results for score differences between ND w/ BC and BC \& ND (n=45)}\label{tb:result_subjective_evaluation_test}
    \begin{tabular}{cccc} \hline
      &ND w/ BC&BC \& ND&p-value \\ \hline
      Hum&5.00&5.14&\phantom{<}.086\phantom{**} \\
      Nat&4.39&4.64&\phantom{<}.019*\phantom{*} \\
      Att&4.87&5.34&<.001** \\ 
      Fac&4.38&4.68&\phantom{<}.012*\phantom{*} \\ \hline
      \multicolumn{4}{r}{*p<0.05  **p<0.01} \\
    \end{tabular}
  \end{center}
\end{table}

Furthermore, Table \ref{tb:result_subjective_evaluation_test} presents the results for statistical tests on the score differences between ND w/ BC and BC \& ND. 
For each metric, a one-tailed paired \textit{t}-test was performed on the distribution of 45 scores obtained for each method.
These results statistically confirm that our proposed method, which predicts various types of nodding, outperforms the conventional method in terms of \textbf{naturalness}, \textbf{attentiveness}, and \textbf{facilitation}.
Most importantly, in the proposed method, backchannel and nodding do not necessarily co-occur. 
In other words, selectively using only backchannel, only nodding, or both, depending on the context, enhances user experience of spoken dialogue systems.

\section{Conclusion} \label{sec:conclusion}
In this paper, we proposed a real-time model that predicts both the timing and types of nodding, one of the nonverbal responses, using audio signals from both the speaker and the listener.
Based on the VAP-based model, we employed multi-task learning with backchannel prediction and VAP pretraining with general dialogue data. 
As a result, we achieved an F1 score of 55.93\% on the timing prediction task and scores of 28.86\%, 39.17\%, and 22.09\% for short, long, and long\_p nodding, respectively, on the timing and type prediction task.
Furthermore, the statistical significance of multi-task learning with backchannel prediction was demonstrated in the timing and type prediction task.
We also evaluated the real-time processing performance and showed that reducing the processing rate to 10 Hz allows the model to operate in real time without a substantial drop in accuracy.

In the subjective experiment, we compared the conventional method which always performs nodding simultaneously with verbal backchannel to our proposed approach, in which backchannel and nodding are predicted separately.
The results revealed that our proposed method received statistically higher ratings in naturalness, attentiveness, and facilitation.

Future work includes developing real-time prediction models for other nonverbal listener responses (e.g., eye gaze and facial expressions) and building multimodal models that leverage the speaker’s gaze and gestures as inputs.
In this study, we deferred incorporating the user's visual cues (e.g., facial expressions) due to real-time computational constraints, and plan to address this by exploring efficient integration methods using a lightweight network.
Additionally, we aim to develop a model for generating listener head motion in real time to create a more natural and responsive avatar.

\section*{Safe and Responsible Innovation Statement}
This study discretely predicts nodding from users’ speech signals. 
Given its nature, we consider the potential for misuse and broader social impact to be minimal.
However, one possible form of misuse would be to falsely claim that a system integrated with the proposed model is operated by a human.
All data used and collected in this study are strictly managed in a local environment. 
We obtained staged consent from participants in the dialogue dataset regarding data release, and the data are recorded in a way that contains no personally identifiable information.
During data collection, gestures were recorded from the same performer who had previously operated an android remotely in another experiment, thus ensuring unbiased gesture data.
The trained model proposed in this study has been released on GitHub, making it accessible for anyone to try.

\section*{Acknowledgements}
This work was supported by JST Moonshot R\&D JPMJPS2011 and JST PRESTO JPMJPR24I4.
In addition, the authors would also like to express their appreciation to Professor Akinobu Lee of Nagoya Institute of Technology for his valuable advice on the software used in this study.

{\footnotesize
\bibliographystyle{unsrt}
\bibliography{sample-base}
}

\end{document}